# Compensation of Fiber Nonlinearities in Digital Coherent Systems Leveraging Long Short-Term Memory Neural Networks

Stavros Deligiannidis, Adonis Bogris, *senior member OSA*, Charis Mesaritakis, Yannis Kopsinis,

*Abstract*—We introduce for the first time the utilization of Long short-term memory (LSTM) neural network architectures for the compensation of fiber nonlinearities in digital coherent systems. We conduct numerical simulations considering either C-band or O-band transmission systems for single channel and multi-channel 16-QAM modulation format with polarization multiplexing. A detailed analysis regarding the effect of the number of hidden units and the length of the word of symbols that trains the LSTM algorithm and corresponds to the considered channel memory is conducted in order to reveal the limits of LSTM based receiver with respect to performance and complexity. The numerical results show that LSTM Neural Networks can be very efficient as post processors of optical receivers which classify data that have undergone non-linear impairments in fiber and provide superior performance compared to digital back propagation, especially in the multi-channel transmission scenario. The complexity analysis shows that LSTM becomes more complex as the number of hidden units and the channel memory increase, however LSTM can be less complex than Digital Back Propagation in long distances (> 1000 km).

*Index Terms*—Fibre nonlinear optics, Optical fibre dispersion, recurrent neural networks, optical transmitters

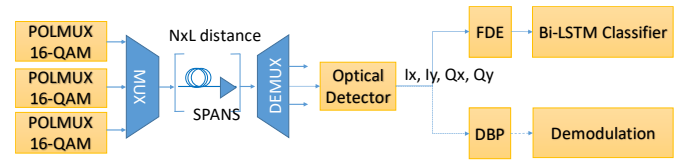

Fig. 1. The simulated transmission link.

## I. Introduction

There is a huge effort in fiber-optic communication industry to cope with the exponentially increasing capacity demands coming from next generation mobile networks and high bandwidth internet applications [1]. New trends such as internet of things especially in the context of tactile internet increase the requirements for real-time, high bandwidth, high availability connectivity in the access network domain, thus enhancing the capacity needs in metro and long-haul transmission networks. Optical fiber communication community predicted the imminent explosion of capacity needs ten years ago and started working intensively on techniques that can leverage fiber capabilities in this respect. After 20 years of high level research there have been identified several solutions involving a combination of advanced modulation formats including probabilistic and/or geometric shaping [2, 3], nonlinear modulation techniques [4] with the use of wavelength division multiplexing (WDM) assisted by space division multiplexing [5] and/or bandwidth extension towards other bands such as O-band [6]. Regardless the capacity enhancement approach that is followed, the major limitation factor of capacity will eventually be the nonlinear Shannon capacity limit of transmitted information. In long-haul high bandwidth optical networks, this limit is mainly attributed to Kerr-induced fiber nonlinearities in their intra-channel and inter-channel form and their interaction with amplified spontaneous emission noise from cascaded optical amplifiers, the so called stochastic parametric noise amplification [7]. Main techniques towards nonlinearity compensation include mid-span optical phase conjugation (OPC) [8], digital back-propagation (DBP) [9], and inverse-Volterra series-transfer function (IVSTF) [10]. OPC is an elegant technique, however it requires broadband wavelength converters and is less compatible to a dynamically configured wavelength routed network. DBP is the most efficient post-processing technique suitable for both linear and nonlinear deterministic effects, since it emulates almost perfectly fiber channel through split-step Fourier with the exception of signal-noise interactions and polarization mode dispersion; however, its real-life implementation still remains impractical due to its high complexity especially when DBP attempts to emulate a multi-channel transmission scenario [11]. IVSTF is a less complex variant compared to DBP, however in principle it is more appropriate for mitigating intra-channel nonlinearity [12]. Lately, there exists an upward trend in the inclusion of machine learning techniques either for the mitigation of transmission

Manuscript received XXXX. This work has been partially funded by the H2020 project NEoteRIC (871330).

Stavros Deligiannidis and Adonis Bogris are with the Department of Informatics and Computer Engineering, University of West Attica, Aghiou Spiridonos, Egaleo, 12243, Athens, Greece (e-mail: sdeligiannid@uniwa.gr, abogris@uniwa.gr).

Charis Mesaritakis is with the Department of Information & Communication Systems Engineering, University of the Aegean, 2 Palama & Gorgyras St., 83200, Karlovassi Samos, Greece (e-mail: cmesar@aegean.gr).

Yannis Kopsinis is with LIBRA MLI Ltd and with LIBRA AI Technologies (ykopsinis@libramli.co.uk).

TABLE I
NUMERICAL MODEL PARAMETERS

| Symbol | Parameter | Value |
|---|---|---|
| G | gain of amplifier | 10dB @ 1550nm , 17dB @1310nm |
| $a$ | attenuation | 0.2dB/km @ 1550nm, 0.34dB/km @1310nm |
| $\beta_2$ | second order dispersion | -0.82 ps$^2$/km @ 1310 nm, -21,5 ps$^2$/km @ 1550 nm ($\lambda_0$=1300 nm) |
| $\gamma$ | fibre nonlinear coefficient | 1.3 W$^{-1}$km$^{-1}$ at both wavelengths |
| R | symbol rate | 25 Gbaud/channel |
| M | modulation format | Dual-polarization 16-QAM |
| L | span distance | 50 km |
| $\Delta f$ | Channel spacing | 50 GHz |

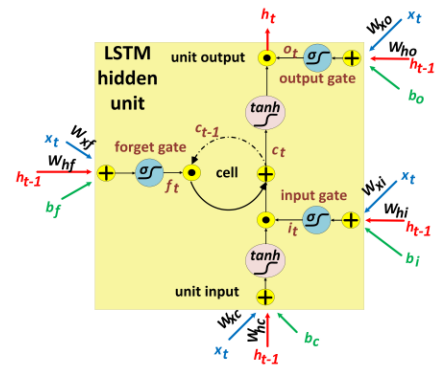

Fig. 2. Conceptual illustration of the LSTM unit

impairments [13] or for the estimation of quality of transmission (QoT) of modern optical communication systems [14]. Different paradigms based on artificial neural networks (ANNs) [15], convolutional neural networks (CNNs) [16], recurrent neural networks (RNNs) [17] are among the techniques that have been successfully applied mostly in intensity modulation/direct detection systems (IM/DD) and in orthogonal frequency division multiplexing (OFDM) [13]. Very recently, bi-directional long short-term memory network (Bi-LSTM) has been proposed for the mitigation of transmission impairments affecting intensity modulation direct detection systems relying on PAM-4 [18, 19] showing that LSTM based signal post-processing can be promising for IM-DD systems. The main advantage of bi-LSTM - and all in principle bi-directional RNNs - is that they can efficiently handle intersymbol interference (ISI) among preceding and succeeding symbols caused by chromatic dispersion. In this paper we examine the efficiency of bi-LTSM in compensating fiber nonlinearity in digital coherent optical communication systems. Although dispersion is compensated very efficiently with the use of linear signal processing at the receiver, its interplay with nonlinearity along the link renders optical fiber a nonlinear channel with memory. Memory size increases with the amount of end-to-end accumulated dispersion. In order to study the efficiency of bi-LSTM based nonlinearity mitigation technique, we carry out numerical simulations at 1550 nm (C-band) and 1310 mn (O-band) considering single-channel and multichannel transmission. As a reference technique for fair comparison, we use DBP. It seems that bi-LSTM outperforms DBP especially in WDM transmission as it can extract and "learn" the nonlinear channel properties that are imprinted as impairments on the signal under specific conditions that would be discussed. The analysis includes the estimation of the bit-error rate (BER) as a function of the number of hidden units and the length of the symbol sequence used to train the network. The analysis also provides complexity estimations and comparison with competing techniques such as DBP. Before referring to basic bi-LSTM features, we first provide a description of the transmission system that is numerically studied in this paper.

## II. TRANSMISSION ANALYSIS

The transmission system depicted in fig. 1 was numerically simulated with the integration of Nonlinear Schrodinger equation (NLSE) using the Split-step Fourier method. We decided to study transmission in two bands, namely the C-band (1550 nm) and the O-band (1310 nm). The reason behind this choice is that C-band remains the dominant band for fiber communications, whilst O-band is expected to be a promising candidate for fiber bandwidth expansion when current C- and L-bands will be fully utilized in the near future. These two bands have significant differences. C-band exhibits the lowest attenuation at high dispersion (0.2 dB/km and 17 ps/nm/km), whilst O-band exhibits almost zero dispersion accompanied by higher losses (0.34 dB/km), thus it is more vulnerable to nonlinearities and their interplay with noise. It is interesting for the system vendors to know whether the same nonlinear equalization technique can be efficient in both environments. We consider single-channel and multi-channel transmission. Each channel is a dual-polarization 16-QAM signal at 25 Gbaud. Lumped amplification was used with span length equal to 50 km and noise figure equal to 5 dB. All parameters are summarized and provided in Table I. The optical receiver depicted in fig. 1 consists of an optical hybrid, balanced photodetectors, low-pass electrical filters with cut-off frequency matched to the baud rate. As fig. 1 shows, we consider two possible post-processing methods, one based on bi-LSTM and one on DBP. When bi-LSTM is used, we first sample the signal and perform ideal chromatic dispersion compensation with the use of an ideal frequency domain equalizer (FDE). When DBP is used, we sample the signal and transfer it to the DBP unit. In this case there is no need for extra dispersion treatment as DBP handles all effects. We assumed ideal carrier phase and frequency estimation (no laser phase noise was taken into account) as well as polarization demultiplexing in both cases as we want to solely focus on nonlinear impairments. Signal propagation in our model is governed by Manakov equations [20].

$$\frac{\partial E_{x,y}}{\partial z} = -\frac{a}{2}E_x + \frac{j\beta_2}{2}\frac{\partial^2 E_{x,y}}{\partial t^2} - j\gamma\frac{8}{9}\left(|E_x|^2 + |E_y|^2\right)E_{x,y} \quad (1)$$

The numerical simulation was performed by integrating the NLSE with the use of split-step Fourier method. $E_x$, $E_y$ are the two orthogonal polarization components of the electric field $E$. We perform the NLSE simulations considering 64 samples per symbol. At the receiver the FDE requires at least 2 samples per symbol in order to provide almost ideal dispersion compensation. DBP operates satisfactorily with 4 samples per symbol. LSTM needs only one sample per symbol for efficient classification, however provided that FDE has cancelled out all dispersion impairments. DBP is applied to a single signal, the channel of interest, that is in our paper DBP is only capable of handling intra-channel effects. Multi-channel DBP is capable of treating inter-channel effects as well, however such an approach enhances significantly the complexity of the DSP at the receiver and thus was not studied at all [11]. Equivalently, bi-LSTM is trained, validated and tested solely based on the information carried by the detected channel of interest. When multi-channel transmission is studied, we study the performance of the central channel in the WDM comb considering that it is the most severely affected by inter-channel effects. Next section describes bi-LSTM implementation.

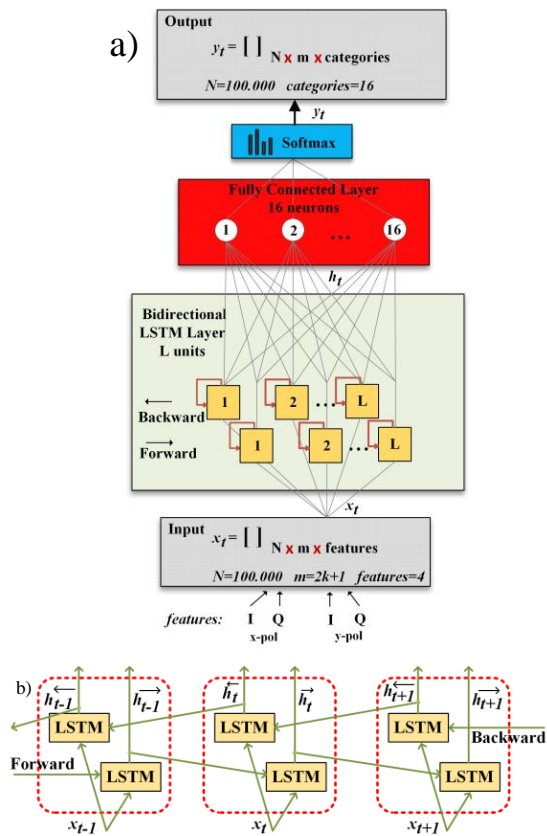

Fig. 3. a) Bidirectional LSTM network architecture, b) Interconnection of adjacent LSTM blocks over time

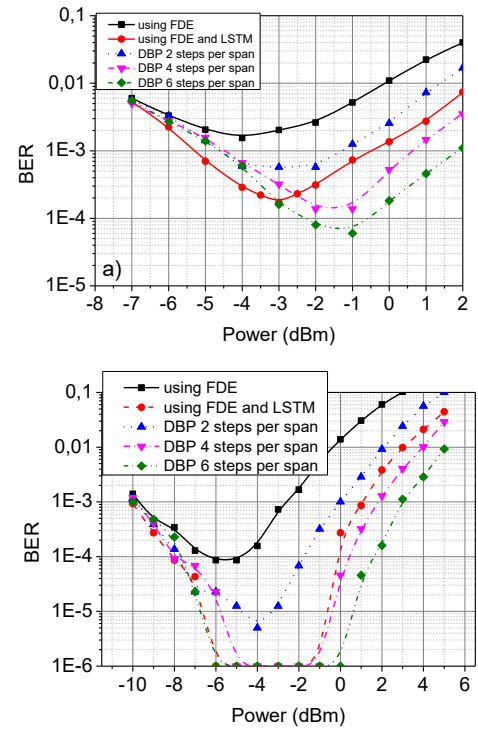

Fig. 4. Single channel transmission a)1310 nm over 700 km. b) 1550 nm over 1500 km.

III. BI-LSTM BASIC FEATURES

RNNs are a type of neural network designed for sequence problems, containing cycles that feed the network activations from a previous time step as inputs to the network to influence predictions at the current time step. These activations are stored in the internal states of the network which can in principle hold long-term temporal contextual information. This mechanism allows RNNs to exploit a dynamically changing contextual window over the input sequence history. The range of contextual information that standard RNNs can access is in practice quite limited. The problem is that the influence of a given input on the hidden layer, and therefore on the network output, either decays or blows up exponentially as it cycles around the network's recurrent connections, referred to as the vanishing gradient problem [21]. LSTM is an RNN architecture specifically designed to address the vanishing gradient problem [22]. Fig. 2 demonstrates an LSTM unit and shows its data flow through input, forget and output gates and a memory cell. The output $h_t$, and cell content $c_t$ are determined by both current input $x_t$ and previous state $h_{t-1}$ under the control of these three gates. The outputs of the gates and content of the cell and state are calculated according to (2).

$$i_t = \sigma(W_{xi}x_t + W_{hi}h_{t-1} + b_i)$$
$$f_t = \sigma(W_{xf}x_t + W_{hf}h_{t-1} + b_f)$$
$$o_t = \sigma(W_{xo}x_t + W_{ho}h_{t-1} + b_o)$$
$$c_t = f_t * c_{t-1} + i_t * tanh(W_{xc}x_t + W_{hc}h_{t-1} + b_c)$$
$$h_t = o_t * tanh(c_t) \quad (2)$$

where $W$ matrices contain the weights of connection $f$, $i$, $o$ and $c$ for forget, input, output gate and cell state respectively. $x_t$, $h_t$, $h_{t-1}$ are input, hidden output, previous hidden output and $b$ are bias vectors. The * operator denotes the element wise product, $\sigma$ is the logistic sigmoid function and $tanh$ is the hyperbolic tangent activation function [23]. In this work we use the sequential neural network demonstrated in Fig 3.

The input $x_t$ is the distorted symbol sequence which has the following form $x_{t,m}=[x_{t-k},…,x_{t-1}, x_t, x_{t+1},…,x_{t+k}]$, where $m$ stands for the overall length of the word which is equal to $m=2k+1$. That is for the symbol at time $t$ we also launch $k$ preceding and $k$ succeeding symbols so as to track intersymbol dependencies. We train the network with the one to one ($m=1$, $k=0$) or many to many approach ($m= 3, 5, 7,…, 201$), with respect to an input sequence $i_{t,m}= =[i_{t-k},…,i_{t-1}, i_t, i_{t+1},…,i_{t+k}]$ that contains the originally sent data and the output $y$ has the same form $y_{t,m}=[y_{t-k},…,y_{t-1}, y_t, y_{t+1},…,y_{t+k}]$. Many to many approach approved to provide better results than many to one. Each symbol in each window contains four values (I and Q for both polarizations) as the input $Xx$-pol and $Xy$-pol feeding the Bidirectional LSTM layer of $L$ hidden units. The many to many approach is beneficial as it takes into account the nonlinear interplay among adjacent bits caused by chromatic dispersion. For larger dispersion values (C-band), the word length must be increased in order to distinguish and "learn" the numerous patterns created as a result of dispersion and nonlinearity interplay. In order to calculate BER of predicted symbols we drive the LSTM network output to a Fully Connected Layer of 16 units and then to a Softmax layer that carries out the classification among 16 classes/QAM symbols of both polarizations for all the symbols at the output $Yx$-pol.

The output at time instance $t$ can be expressed as:

$$y_t = Softmax[W_{Softmax}h_t + b_{Softmax}] \quad (3)$$

where $h_t = \frac{1}{2}(\overrightarrow{h_t} + \overleftarrow{h_t})$ and $\overrightarrow{h_t}$ and $\overleftarrow{h_t}$ are the outputs of the LSTM in the forward and the backward directions respectively. Softmax is defined as $y=Softmax(z)$ and is an activation function that assigns decimal probabilities to each class in a multi-class problem with $y_i = \frac{e^{z_i}}{\sum_{16} e^{z_i}}$ in our case. BER is calculated only for the central symbol of each $y_{t,m}$ word with the respect to its $i_{t,m}$ counterpart, although our simulations showed that many symbols within $y_{t,m}$ word could be successfully detected at the same run.

The LSTM network is built, trained and evaluated in Keras with Tensorflow 2.1.0 GPU backend. In the Keras model, categorical cross-entropy error is chosen as a loss function and Adam as the optimizer for the BER measurement with the parameters appearing in [24]. We consider 60% for training, 20% for validation and 20% for testing with unknown data. The training stage is executed with batches of 512 symbols for optimum balance between memory allocation size and execution time. The maximum forward and backward passes of all the training symbols (epochs) is chosen to be 400. To avoid overfitting during training we use "early stopping" when validation accuracy does not improve for 20 successive epochs.

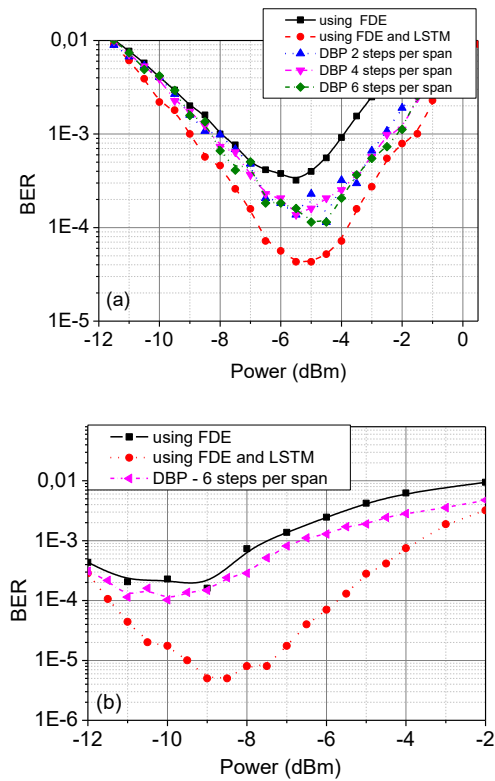

Fig. 5. 10 WDM channel transmission a) L=300 km, λ=1310 nm b) L=500km, λ=1550 nm

## IV. RESULTS AND DISCUSSION

### A. Transmission results in single-channel and multi-channel operation

Transmission simulations were performed in Matlab for single channel transmission and at a second stage for multi-channel transmission/compensation considering up to 10 WDM channels at 50 GHz spacing. We carefully selected the process for providing pseudorandom unrepeated sequences using rng('shuffle') and the very long period of $2^{19937}-1$ Mersenne Twister generator, so as to avoid bi-LSTM to predict the next symbol of the pseudorandom sequence and overestimate the nonlinearity mitigation results [25]. We have used 60000 symbols for training, 20000 symbols for validation and 20000 symbols for testing. In order to estimate very low BER values, (see fig. 4), we add extra batches of 20000 unknown symbols in the BER estimation. The lowest possible BER that could be numerically counted is $10^{-6}$. Although single-wavelength transmission is not the case for long-haul transmission systems, the efficacy of a compensation technique targeting fiber nonlinearities must be first assessed when self-phase modulation is the dominant effect (intra-channel nonlinearity). Fig. 4 shows the performance for different post-processing scenarios, namely FDE only equalization, FDE followed by bi-LSTM, DBP employing 2 up to 6 steps per span at 1310 nm (a) and 1550 nm (b). Bi-LSTM in this series of calculations comprises $L=32$ hidden units in the LSTM layer and a word length equal to $m=71$ symbols for both 1310 nm and 1550 nm. Regarding DBP, it is obvious that as the number of steps per span increase, DBP becomes more accurate at expense of complexity. Fig. 4 shows that 1550 nm is much more tolerant to nonlinear effects, as expected, due to lower attenuation and higher dispersion. Bi-LSTM has the ability to improve BER more than an order of magnitude compared to typical linear equalization especially in the C-band (more than two orders of magnitude improvement) and outperforms DBP employing 2 steps per span, performs equally with DBP using 4 steps per span whilst it is less efficient compared to DBP employing 6 steps per span. It must be reminded that DBP in our simulations require 4 samples per symbol, whilst LSTM based classification operates efficiently with only one sample per symbol.

We continued our analysis in the multi-channel regime. In this case we focused on the central channel and for all post-processing methods we only take into account the information provided by the detected channel. This choice is not the optimal, as information from neighboring channels would contribute to deeper knowledge of the channel; this choice reflects the need for proposing a receiver of moderate complexity [11]. The results for both bands are depicted in fig. 5. In the multi-channel transmission, bi-LSTM proves to be superior to DBP as the latter equalizes solely intra-channel effects and ignores inter-channel ones. Although bi-LSTM is trained based on the information provided by the central channel as well, it seems that it has the ability to adequately capture inter-channel effects of the nonlinear medium that affect the detected channel and provide classification with much better results when compared to FDE and DBP. The possibility of identifying channel information even in single-user detection has been discussed in the past [26 and references there in] and is based on frequency-resolved logarithmic perturbation which shows that optical systems can be modeled as linear time-varying systems, where the effect of inter-channel nonlinearity emerges as linear time-varying inter-symbol interference. If the coherence time of the channel is much longer than symbol period [26], the XPM effects are sufficiently slow to be "deterministically" tracked by the LSTM equalizer. In the O-band, bi-LSTM manages to improve the BER by an order of magnitude compared to FDE and half-order of magnitude compared to DBP. Its efficacy is pronounced in the C-band, where the provided BER improvement surpasses an order of magnitude compared to FDE, DBP. The main reason for better results in the C-band is that inter-channel effects in this case are dominated by slow XPM variations, whilst in the O-band, the strength of inter-channel effects increases as a result of lower dispersion. O-band enhances the strong presence of four-wave mixing (FWM) between WDM signals and between signal and noise [7].

Although we have already shown that bi-LSTM can be efficient in handling nonlinear impairments, there has been no analysis on the impact of LSTM internal properties on the classification performance. The role of the number of hidden layers and the length of the symbol word is highlighted in fig. 6. It becomes evident that as the number of hidden units increases, the BER performance improves. We can safely claim that for a number of hidden units exceeding 20, the BER performance reaches the optimal value. The impact of the word length is also very significant and relates to channel memory determined by accumulated dispersion. In order to better highlight the underlying reasons of this behavior it must be noted that we accomplish classification with the use of the softmax layer intending to directly match the output of the LSTM network $y_t$ to the original transmitted 16-QAM symbol per polarization $x_t$ given the word that contains $k$ preceding and $k$ succeeding symbols of $x_t$, that is $x_{t,m}=[x_{t-k},…,x_{t-1}, x_t, x_{t+1},…,x_{t+k}]$, where $m$ stands for the overall length of the word which is equal to $m=2k+1$. The longer the sequence that is used in order to train the network in the many to many approach, the larger the memory of channel that is considered before LSTM processing. It is well known that fiber channel memory is linked to the delay spread caused by chromatic dispersion. When a signal of optical bandwidth equal to $B$ propagates along a link of length $L$ it experiences a delay spread equal to $\Delta t_{CD}=2\pi|\beta_2|LB$. At 1310 nm this simple formula gives a delay spread of 77 ps, thus the intra-channel ISI extends to a duration of two pulses (25 Gbaud corresponds to 40 ps pulse duration). On the contrary, at 1550 nm the delay spread amounts to 3.4 ns for 500 km, which means that the intra-channel nonlinear effects could involve ISI among 85 adjacent symbols. Taking into account WDM transmission, it is well known that the channel memory increases due to walk-off among co-propagating waves, however the nonlinear interaction reduces as a result of the same effect. Fig. 6 clearly shows that in the O-band, where the memory of the channel is shorter (2 symbols in the single-channel case), the BER results reach their optimal value for a word length above 3 symbols. On the contrary, in the C-band where the finite memory of the channel increases due to higher dispersion, the optimal value is reached for a word length surpassing 50 symbols.

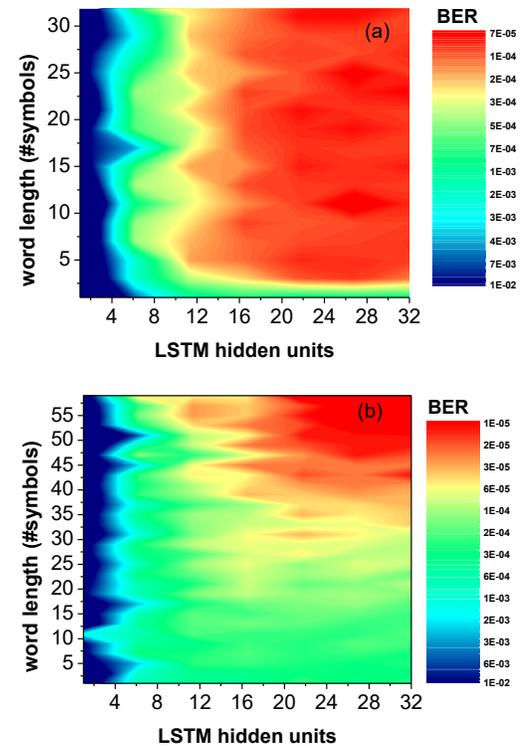

Fig. 6. BER as a function of LSTM hidden units and the number of symbols per word in the many to many approach. a) 1310 nm, 300 km, 10 WDM channels, b) 1550 nm, 500 km, 10 WDM channels

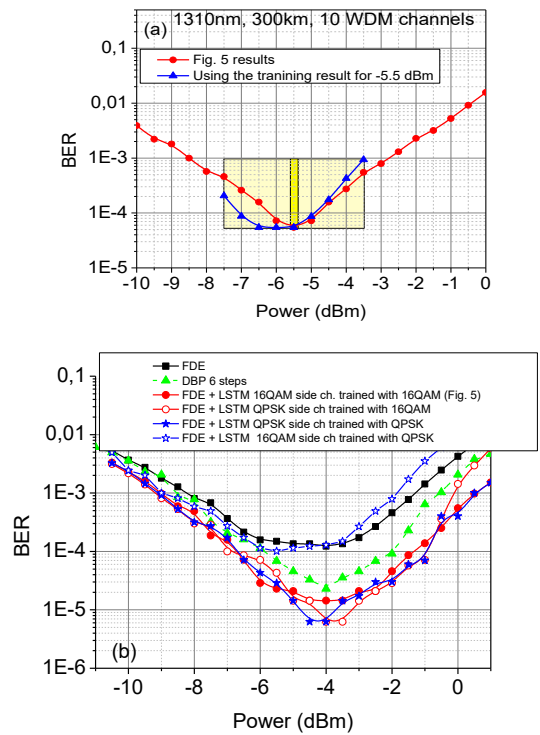

Fig. 7. Evaluation of LSTM training robustness to channel variations: a) Sensitivity of training to power variations (1310 nm, 16-QAM, 300 km), b) Sensitivity of training to the modification of modulation format of neighboring channels (1310 nm, QPSK neighbors for a 16-QAM central wavelength, 300 km).

### B. Robustness of training in channel variations

The next subject of this paper is to identify the robustness of bi-LSTM training against small variations of the channel. In general, fiber channel is less time-varying when compared to wireless channel. The main variations that could affect the stability of the transmission link are power variations, polarization rotations and the change of the modulation formats of neighboring WDM channels in dynamic optical networks. Another critical parameter is the change of transmission length, however this can be handled if the receiver is a priori trained for different transmission distances. Polarization rotations occur at relatively low rates and are handled by proper algorithms in the DSP unit of the optical receiver. In this paper, we consider as most important changes those that refer to power variations and variable modulation formats of neighboring channels as they directly affect the nonlinear properties of the channel. Power affects nonlinearity through Kerr effect and inter-channel impairments depend on neighboring channels power and modulation format. For instance, it is well known that XPM effect depends on the fourth order moment of the modulation alphabet of the adjacent channels [26]. In fig. 7 we study the robustness of training with respect to power and modulation format variations. In fig. 7a, the BER at the output of the O-band WDM transmission is depicted for the central channel. The red line with circles shows the result of fig. 5 which refer to the performance obtained with the use of bi-LSTM after performing training for each power value shown in the figure. After having completed this process, we observe that optimum performance is obtained at -5.5 dBm average power per channel. Thus, if we consider that the transmission link remains unaltered, the receiver should be operated based on the training obtained at -5.5 dBm. Then, the question is what the performance will be if the average power is moderately changed due to unexpected events in the link (i.e. the gain of amplifiers is changed, the power of the transmitter is reduced, etc.). The blue line in fig. 7a answers this question as it shows how transmission will perform if the receiver is fixed in the training obtained at -5.5 dBm. For power values ranging from -7.5 dBm to -3.5 dBm (+/- 2dB variation around the nominal value of -5.5 dBm) the BER is not substantially affected. On the contrary, it seems that if one applies training knowledge obtained at -5.5 dBm at lower power values, the BER will be slightly improved, thus the training obtained at the optimum point acts beneficially at lower power values. The main explanation for this behavior is that at the point the system performs optimal classification, there is a trade-off between linear and nonlinear noise. The obtained training can be applied at lower power values, as the nonlinear noise will be less, thus a system which has learnt to operate in a harsher nonlinear environment can adapt to a less nonlinear ecosystem.

The second way to test the robustness of the obtained training against variations of transmission channel was to change the modulation format of adjacent WDM channels and re-evaluate the performance for the initially obtained training results. Thus, we trained the bi-LSTM considering the following scenarios in fig. 7b): i) (full-red dot) network trained with 16-QAM neighboring channels and tested on 16-QAM neighboring channels, ii) (empty-red dot) network trained with 16-QAM neighboring channels but tested on QPSK neighboring channels, iii) (full-blue star) network trained with QPSK neighboring channels and tested on QPSK neighboring channels and iv) (empty-blue star) network trained with QPSK neighboring channels but tested on 16-QAM neighboring channels. Our aim was to see whether the weights of the neural network obtained for a purely POLMUX 16-QAM transmission could be valid for a transmission system where the central channel remains 16-QAM and neighboring channels revert to POLMUX-QPSK and vice versa. Based on the results of fig. 7b, it seems that the receiver based on training knowledge obtained in the case of 16-QAM transmission provides more or less the same BER performance with the receiver that was particularly trained in a transmission system where the central channel is 16-QAM and all the other wavelengths carry QPSK signals. On the contrary, when the network is trained with QPSK neighboring channels but tested on 16-QAM neighboring channels the performance degradation is obvious, showing the weakness of LSTM network to mitigate the signal in harsher environments (16-QAM neighbors) than the one it has been trained on (QPSK neighbors). Therefore, bi-LSTM training is robust against modulation format variations of neighboring channels provided that training has taken place with the most complex modulation format foreseen to be applied in the system. In this respect, the receiver is compatible with next generation optical networks performing frequent modulation format modifications with the use of software defined transceivers depending on dynamic network capacity needs.

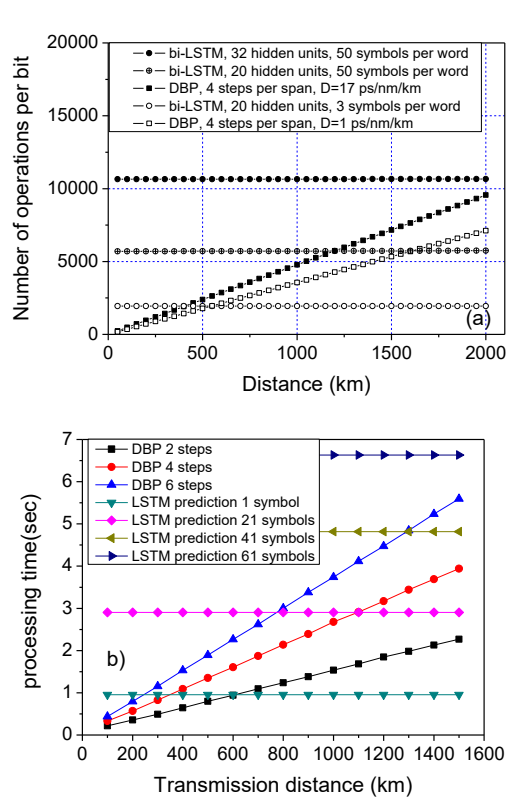

Fig. 8. a) Analytically calculated computational complexity for bi-LSTM with FDE vs DBP. b) Estimated processing time when both algorithms were simulated in the GPU for diverse operating conditions, H=32 in this case.

### C. Complexity analysis

This section focuses on the hardware complexity of bi-LSTM and how it compares with DBP. In order to perform this comparison, we estimated the amount of multiplications per bit needed by each approach. For DBP algorithm the complexity per bit can be evaluated according to [27]

$$C_{DBP} = 4N_{Span}N_{StpSt}\left[\frac{N(log_2 N+1)n_s}{(N-N_D+1)log_2 M} + n_s\right] \quad (4)$$

where $N_{Span}$ is the number of spans, $N_{StpSt}$ is the number of steps per span, $N$ is the FFT size (depending on accumulated dispersion per span), $n_s$ is the oversampling factor, $M$ is the constellation order and $N_D = \frac{n_s \tau_D}{T}$, where $\tau_D$ corresponds to the dispersive channel impulse response and $T$ is the symbol duration. We multiply the complexity by 4 since for DBP we use complex numbers and one complex multiplication is equal to four real multiplications. For the bi-LSTM implementation we must add the complexity of the frequency domain equalizer [27]

$$C_{FDE} = 4 \frac{N(log_2 N+1)n_s}{(N-N_D+1)log_2 M} \quad (5)$$

Here $N$ corresponds to FFT size taking into account the accumulated dispersion along the entire link. The complexity of bi-LSTM architecture consists of two parts: the training (6) and the prediction complexity (7). The computational complexity of the proposed bi-LSTM network depends on the number of hidden units and the number of inputs (length of the symbol sequence) [18, 28, 29]

$$C_{train} = N_{ep}N_{TS}C_{pred} \quad (6)$$

$$C_{pred} \sim 16(L^2 + Lm + L)/log_2 M \quad (7)$$

where $N_{ep}$ is the number of training epochs, $N_{TS}$ the number of symbols used for training, $L$ the number of LSTM hidden units (fig. 3b) and $m=2k+1$ the number of input/output symbols respectively. In order to calculate the complexity for bi-LSTM one has to identify how often training is expected to take place. Although this information can be safely extracted in live experiments, one could assume that since bi-LSTM is robust to power and modulation format changes, the only time-varying effect that is anticipated to disturb the channel is random polarization rotation. The time scales for polarization dynamics exceed a few ms. Within a ms, almost $5 \times 10^7$ will have been transmitted per polarization, whilst training requires about 20000 symbols for 100 epochs, which is equivalent to a time period of $2 \times 10^6$ symbols. Therefore, even if the optical channel varies every ms, the training complexity is less than 4% of the prediction complexity and thus it can be safely ignored. Having this in mind, in fig 8a we calculate the number of operations per bit as a function of transmission distance for DBP and LSTM taking into account only prediction complexity for bi-LSTM and FDE complexity on top of that. As can be seen in fig. 8, DBP complexity grows linearly as a result of the number of spans appearing in (3). Moreover, DBP is not so much dependent on the chromatic dispersion of the span, that is the complexity for a specific distance is not that much different in the O-band and in the C-band. It must be reminded that as the dispersion reduces, the parameter $N$, that is the FFT size varies proportionally (typical values 64, 128, 256 for each span) and this is the reason for complexity changes. In contrast, LSTM has a high complexity which however is not that much dependent on the transmission distance if one considers constant word length in the training process (with the exception of FDE complexity which is however small). Moreover, its complexity vastly reduces when the number of symbols in the input/output word reduces and this depends on the anticipated accumulated dispersion which depends on transmission distance. That means that LSTM could be very efficient in a nonlinear channel with small memory which is the case of O-band or C-band with the use of dispersion management. For 20 hidden units and C-band transmission, LSTM with word length=50 becomes less complex than DBP for distances surpassing 1200 km, whilst the same tendency is observed below 500 km when transmission takes place in the O-band (word length=3). In order to obtain a more realistic picture of these trends, we conducted numerical simulations of both algorithms in a GPU GTX 1070. DBP was simulated in Matlab with GPU Coder using CUDA 10.1 toolkit development environment that includes GPU-accelerated libraries, debugging and optimization tools, a C/C++ compiler and a runtime library. Bi-LSTM was trained and tested in Tensorflow 2.1.0 with the same GPU using the same CUDA 10.1 Toolkit. For the same number of symbols (20000) we observed similar trends that clearly show the dependence of LSTM complexity on the number of hidden units and symbols and its tendency to operate more efficiently than DBP at distances surpassing 1000 km.

## V. CONCLUSIONS

This paper showed the potential of bi-LSTM neural networks as post-processing engines in digital coherent systems. LSTM has similar efficiency with DBP in mitigating intra-channel effects and superior performance when inter-channel effects are present. The analysis showed how the performance of the algorithm is affected by important parameters (number of hidden units, number of symbols at input/output) and showed that training is tolerant to power variations and modifications of neighboring channels modulation provided that training has been accomplished in the worst nonlinear environment expected. Finally, the complexity analysis showed that LSTM could compete DBP especially in long distances and in optical channels exhibiting small accumulated dispersion where LSTM complexity can be reduced.

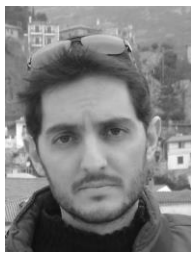
**Stavros Deligiannidis** (holds a BSc in Physics, a MSc degree in Microelectronics and VLSI from the National and Kapodistrian University of Athens. Since 2010 he is with the Department of Computer Engineering of the Technological Educational Institute of Peloponesse, Greece where he serves as a Lecturer. He is currently pursuing his PhD degree at the University of West Attica in the field of novel signal processing techniques for optical communication systems under the supervision of Prof. Adonis Bogris. He has worked as a researcher in local and European projects. His current research interests include optical communications, deep learning, digital signal processing, and parallel computing.

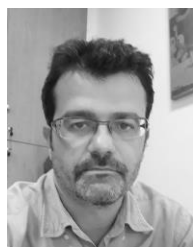
**Adonis Bogris** was born in Athens. He received the B.S. degree in informatics, the M.Sc. degree in telecommunications, and the Ph.D. degree from the National and Kapodistrian University of Athens, Athens, in 1997, 1999, and 2005, respectively. His doctoral thesis was on all-optical processing by means of fiber-based devices. He is currently a Professor at the Department of Informatics and Computer Engineering at the University of West Attica, Greece. He has authored or co-authored more than 150 articles published in international scientific journals and conference proceedings and he has participated in plethora of EU and national research projects. His current research interests include high-speed all-optical transmission systems and networks, nonlinear effects in optical fibers, all-optical signal processing, neuromorphic photonics, mid-infrared photonics and cryptography at the physical layer. Dr. Bogris serves as a reviewer for the journals of the IEEE and OSA.

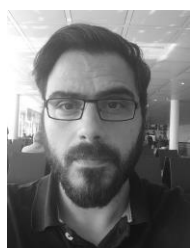
**Charis Mesaritakis** received his BS degree in Informatics, from the department of Informatics & Telecommunications of the National & Kapodistrian University of Athens in 2004. He received the MSc in Microelectronics from the same department, whereas in 2011 he received his Ph.D degree on the field of quantum dot devices and systems for next generation optical networks, in the photonics technology & optical communication laboratory of the same institution. In 2012 he was awarded a European scholarship for post-doctoral studies (Marie Curie FP7-PEOPLE IEF) in the joint research facilities of Alcatel-Thales-Lucent in Paris-France, where he worked on intra-satellite communications. He has actively participated as research engineer/technical supervisor in more than 10 EU-funded research programs (FP6-FP7-H2020) targeting excellence in the field of photonic neuromorphic computing, cyber-physical security and photonic integration. He is currently an Associate Professor at the Department of Information & Communication Systems Engineering at the University of the Aegean, Greece. He is the author and co-author of more than 60 papers in highly cited peer reviewed international journals and conferences, two international book chapters, whereas he serves as a regular reviewer for IEEE and OSA.

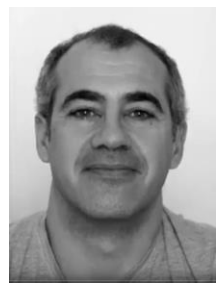
**Yannis Kopsinis** has received his Ph.D. from the Dept. of Informatics and Telecommunications, Univ. of Athens in 2004. Since then he has gained a number of prestigious personal research grants. Among them, a Marie Curie IEF fellowship and a Ramón y Cajal Fellowship, University of Granada, Spain. Moreover, he has worked for more than 5 years as a senior research fellow in the School of Engineering and Electronics, the University of Edinburgh, UK. He has published more than 50 papers in technical journals and conferences and he has co-authored 3 book chapters. He has worked on Signal Processing & Machine Learning theory and methods for more than 2 decades in applications ranging from telecoms to audio and medical. His current research interests lie in the areas of Online Machine Learning, deep learning, cooperative & federated learning, constrained optimization and robust predictive analytics. Since 2015 he is the director of LIBRA MLI Ltd and since 2020 he is the CEO of LIBRA AI Technologies IKE.